%% file: z-was.tex
\documentclass[12pt]{article}
\usepackage{epsfig}

\newcommand\pubnumber{UTHEP-01-0103}
\newcommand\pubdate{\today}
\newcommand\hepnumber{hep-ph/0101246}

\def\support{\footnote{Work supported in part by 
        the US DoE contracts DE-FG05-91ER40627 and DE-AC03-76SF00515,
        the European Comission 5-th framework contract HPRN-CT-2000-00149
        and the Polish--French Collaboration within IN2P3 through LAPP Annecy.}}

\def\Title#1{\begin{center} {\Large\bf #1 } \end{center}}
\def\Author#1{\begin{center}{ \sc #1} \end{center}}

\newcommand\pubblock{\rightline{\begin{tabular}{l} \pubnumber\\
         \pubdate\\ \hepnumber \end{tabular}}}
\newenvironment{Abstract}{\begin{quotation}  }{\end{quotation}}
\newenvironment{Presented}{\begin{quotation} \begin{center} 
             Presented at the\end{center}
      \begin{center}\begin{large}}{\end{large}\end{center} \end{quotation}}

\makeatletter
\def\section{\@startsection{section}{0}{\z@}{5.5ex plus .5ex minus
 1.5ex}{2.3ex plus .2ex}{\large\bf}}
\def\subsection{\@startsection{subsection}{1}{\z@}{3.5ex plus .5ex minus
 1.5ex}{1.3ex plus .2ex}{\normalsize\bf}}
\def\subsubsection{\@startsection{subsubsection}{2}{\z@}{-3.5ex plus
-1ex minus  -.2ex}{2.3ex plus .2ex}{\normalsize\sl}}

\renewcommand{\@makecaption}[2]{%
   \vskip 10pt
   \setbox\@tempboxa\hbox{\small #1: #2}
   \ifdim \wd\@tempboxa >\hsize     
       \small #1: #2\par          
     \else                        
       \hbox to\hsize{\hfil\box\@tempboxa\hfil}
   \fi}

 \def\citenum#1{{\def\@cite##1##2{##1}\cite{#1}}}
 
\newcount\@tempcntc
\def\@citex[#1]#2{\if@filesw\immediate\write\@auxout{\string\citation{#2}}\fi
  \@tempcnta\z@\@tempcntb\m@ne\def\@citea{}\@cite{\@for\@citeb:=#2\do
    {\@ifundefined
       {b@\@citeb}{\@citeo\@tempcntb\m@ne\@citea\def\@citea{,}{\bf ?}\@warning
       {Citation `\@citeb' on page \thepage \space undefined}}%
    {\setbox\z@\hbox{\global\@tempcntc0\csname b@\@citeb\endcsname\relax}%
     \ifnum\@tempcntc=\z@ \@citeo\@tempcntb\m@ne
       \@citea\def\@citea{,}\hbox{\csname b@\@citeb\endcsname}%
     \else
      \advance\@tempcntb\@ne
      \ifnum\@tempcntb=\@tempcntc
      \else\advance\@tempcntb\m@ne\@citeo
      \@tempcnta\@tempcntc\@tempcntb\@tempcntc\fi\fi}}\@citeo}{#1}}
\def\@citeo{\ifnum\@tempcnta>\@tempcntb\else\@citea\def\@citea{,}%
  \ifnum\@tempcnta=\@tempcntb\the\@tempcnta\else
  {\advance\@tempcnta\@ne\ifnum\@tempcnta=\@tempcntb \else\def\@citea{--}\fi
    \advance\@tempcnta\m@ne\the\@tempcnta\@citea\the\@tempcntb}\fi\fi}
\makeatother

\def\Order#1{${\cal O}(#1$)}

\def\Order#1{${\cal O}(#1)$}

\def\KK{${\cal KK}$}

\def\bbeta{\bar{\beta}}
\def\hbeta{\hat{\beta}}
\newcommand{\sfac}{\mathfrak{s}}
\usepackage{amssymb}
\usepackage{euscript}
\newcommand{\Meu}{\EuScript{M}}

\input econfmacros2.tex

\begin{document}
\begin{titlepage}
\pubblock

\vfill
\def\thefootnote{\fnsymbol{footnote}}

\Title{ Coherent Exclusive Exponentiation  of 2f Processes  in $e^+e^-$ Annihilation\support}
\vfill
\Author{Z. W\c{a}s,%
        \footnote{Institute of Nuclear Physics, Cracow, ul. Kawiory 26A, Poland}
        S. Jadach%
        \footnote{Theory Group DESY, Platanenallee 6, D-15738 Zeuthen, Germany \\
                 and
                 Institute of Nuclear Physics, Cracow, ul. Kawiory 26A, Poland}
        {\it and}
       B.F.L. Ward%
        \footnote{Department of Physics and Astronomy,
                 The University of Tennessee, Knoxville, Tennessee 37996-1200, USA},
}

\vfill
\begin{Abstract}
In the talk we present the Coherent Exclusive Exponentiation (CEEX)
which is implemented in the \KK MC event generator
for the process $e^+e^-\to f\bar{f} +n\gamma$, $f=\mu,\tau,d,u,s,c,b$
for center of mass energies from $\tau$ lepton threshold to 1TeV,
that is for LEP1, LEP2, SLC, future Linear Colliders, $b,c,\tau$-factories etc.
We will attempt a short discussion of the theoretical concepts necessary 
in our approach, in particular the relations between the rigorous calculation 
of spin amplitudes (perturbation expansion), phase space parametrisation and 
exponentiation. 
In CEEX effects due to photon emission from the initial beams and outgoing fermions
are calculated in QED up to second-order, including all interference effects.
Electroweak corrections are included in first-order, at the amplitude level.
The beams can be polarised longitudinally and transversely,
and all spin correlations are incorporated in an exact manner.
Precision  predictions, in particular 
the photon emission at LEP2 energies, are also shown.
\end{Abstract}
\vfill
\begin{Presented}
5th International Symposium on Radiative Corrections \\ 
(RADCOR--2000) \\[4pt]
Carmel CA, USA, 11--15 September, 2000
\end{Presented}
\vfill
\end{titlepage}
\def\thefootnote{\arabic{footnote}}
\setcounter{footnote}{0}
%

\hyphenation{author another created financial paper re-commend-ed}


\section{Introduction}
At the end of LEP2 operation the total cross section for 
the process $e^-e^+\to f\bar{f}+n\gamma$
has to be calculated with the precision $0.2\%-1\%$, depending on the event 
selection.
The arbitrary differential distributions have to be calculated 
with the corresponding precision.
Even now, this is not always the case \cite{Kobel:2000aw} and the calculations 
are still continuing.
In the future, for linear colliders (LC's), the precision requirement can be even more demanding.
These requirements necessitate development of the appropriate 
calculational schemes 
for the QED corrections and the construction of new dedicated MC programs.
We present here an effort in this direction.
Our report is based on refs.~\cite{ceex1:1999,gps:1998,ceex2:2000}
and the Monte Carlo program is described in ref.~\cite{kkcpc:1999}.
The pedagogical introduction to some concepts necessary in understanding 
exponentiation can be found e.g. in~\cite{Was:1994kg}.

\begin{table*}[!ht]
\centering
\setlength{\unitlength}{0.1mm}
\begin{picture}(1400,900)
\put( 0, 0){\makebox(0,0)[lb]{\epsfig{file=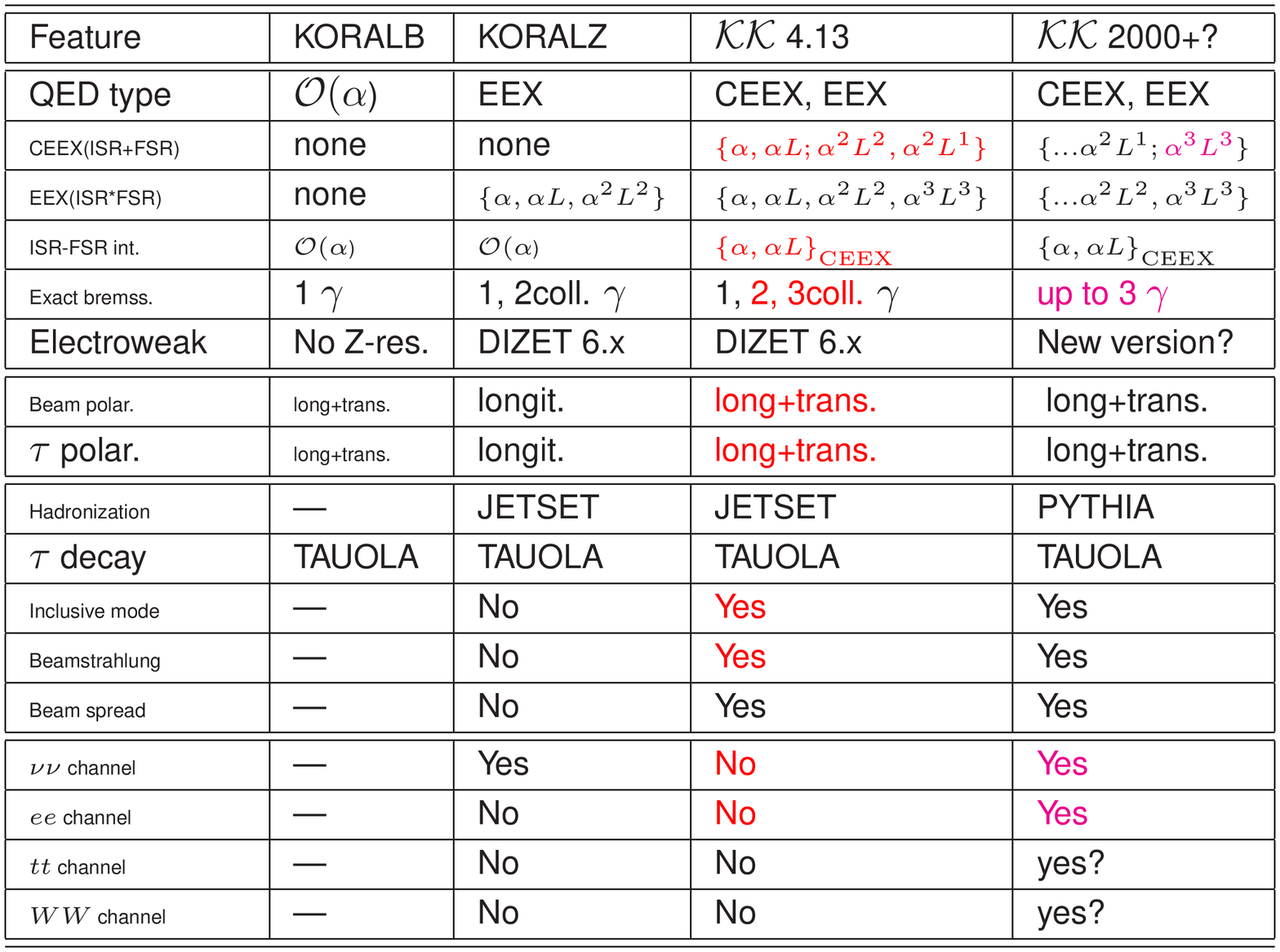,width=140mm,height=90mm}}}
\end{picture}
\caption{Overview of the \KK MC event generator as compared with KORALZ and KORALB.
}
\label{tab:compare}
\end{table*}

\section{What is precision calculation}
New results from high energy particle experiments are obtained as a result of 
the huge effort of hundreds of experimental physicists over
many years. In the cases when theoretical calculations are needed to interpret 
the  results, it is fair to require, whenever possible, 
the uncertainty of the calculations
to be smaller at least by a factor of 3  than the experimental error.
Once the condition is fulfilled, in the final interpretation of 
experimental data for quantities such as coupling constants, 
total cross sections or particle masses, 
the final combined theoretical and experimental uncertainty would not increase more
than 10 \%  with  respect to experimental uncertainty alone.
This rule of thumb is  motivated in cases when the theoretical calculations
are possible and  
require an effort much smaller than that of the experiments. 

The crucial requirement of the high precision calculation is however not only
that its results agree with the measured data, but also, that the relation
of the results with the foundation of the Standard Model field theory 
is fully controlled. At present, requirements for precision, as 
defined by experiments, do not exceed the 0.1 \% tag. That is why, in general,
predictions including complete Standard Mode corrections of \Order{\alpha_{QED}},
 are sufficient.  Only those terms of the higher orders which include 
enhancement factors such as $\ln{s \over m_f^2}$,  $\ln{M_Z \over \Gamma_Z}$, $m_t \over m_W$ etc., have to be taken into account. 

Thanks to this, one can define schemes of calculation where QED calculations
can be separated from the rest, and dealt with to large degree individually%
\footnote{Even further separation is possible: emission of additional real
fermion pairs can be calculated separately. At the same step,
the appropriate virtual corrections have to be included into predictions for
$2f$-processes.}.
As it was presented in  \cite{Kobel:2000aw,Grunewald:2000ju} this was 
indeed the solution successful for LEP2  
$e^+e^- \to 2f$ and  $e^+e^- \to 4f$ processes.
Exponentiation is a convenient way of dealing with the QED corrections,
which are large, and depend on the detection conditions (cuts).
\section{What is coherent exclusive exponentiation CEEX?}
The {\em exponentiation} is generally a method of summing up real and virtual photon
contributions to infinite order such that infrared (IR) divergences cancel.
The {\em exclusivity} means that the procedure of exponentiation, that is summing up
the infrared (IR) real and virtual contribution, within the standard perturbative scheme
of quantum field theory, is done at the level of the fully differential (multiphoton) cross section,
or even better, at the level of the scattering matrix element (spin amplitude),
{\em before any phase-space integration over photon momenta is done}.
The other popular type of the exponentiation is {\em inclusive} exponentiation (IEX), 
which is done at the level of inclusive distributions, structure functions, etc.
see discussion in ref.~\cite{sussex:1989}.
The classical work of  Yennie-Frautschi-Suura~\cite{yfs:1961} (YFS) represents the best
example of the exclusive exponentiation and we nickname it as EEX.
Finally, why do we use word {\em coherent?} In CEEX the
essential part of the summation of the IR real and virtual photon contributions
is done at the amplitude level.
Of course, IR cancellations occur as usual at the probability level, however,
the transition from spin amplitudes to differential cross sections,
and the phase space integration are done entirely numerically!
As a consequence of the above {\em coherent} approach it follows
that CEEX is friendly
to coherence among Feynman diagrams, narrow resonances, interferences, etc.
This is a great practical advantage.
In our many previous works which led to the development of the Monte Carlo
event generators like YFS2, YFS3, KORALZ, KORALW, YFS3WW, BHLUMI, BHWIDE,
see refs.~\cite{yfs2:1990,koralz4:1994,koralw:1998,yfsww:1996,bhlumi4:1996,bhwide:1997},
we have generally employed EEX, which is closely related to the YFS work~\cite{sussex:1989}.
The CEEX is a recent development and is so far used only in 
the new \KK MC program~\cite{kkcpc:1999}.

Let us now show in a very simplified schematic way what is the the main difference
between the old EEX/YFS and the CEEX for the fermion pair production the process:
\begin{equation}
 e^-(p_1,\lambda_1)+e^+(p_2,\lambda_2) \to 
 f(q_1,\lambda'_1)+\bar{f}(q_2,\lambda'_2)+\gamma(k_1,\sigma_1)+...+\gamma(k_n,\sigma_n).
\end{equation}
The EEX total cross section is
\begin{equation}
\sigma = \sum\limits_{n=0}^\infty \;\;
          \int\limits_{m_\gamma}
          d\Phi_{n+2}\; e^{Y({m_\gamma})}
          D_n(q_1,q_2,k_1,...,k_n),
\end{equation}
where in the \Order{\alpha^1} the distributions for $n_\gamma=0,1,2$ are
\begin{eqnarray}
    D_0          =&  \bbeta_0 \nonumber \\
    D_1(k_1)     =&  \bbeta_0 \tilde{S}(k_1) +\bbeta_1(k_1)  \nonumber \\
    D_2(k_1,k_2) =&  \bbeta_0 \tilde{S}(k_1)\tilde{S}(k_2) 
                  +\bbeta_1(k_1)\tilde{S}(k_2)+\bbeta_1(k_2)\tilde{S}(k_1)
\end{eqnarray}
and the real soft factors are defined as usual
\begin{equation}
 4\pi\tilde{S}(k) = \sum\limits_\sigma |\sfac_\sigma(k)|^2 = |\sfac_+|^2(k) +|\sfac_-(k)|^2\\
             = -{\alpha\over \pi}\bigg({q_1\over kq_1}-{q_2\over kq_2} \bigg)^2.
\end{equation}
What is important for our discussion is that the IR-finite building blocks 
\begin{eqnarray}
  \bbeta_0=& \sum_\lambda |\Meu_\lambda|^2, \nonumber\\
  \bbeta_1(k)=&  \sum\limits_{\lambda\sigma} |\Meu^{\rm 1-phot}_{\lambda\sigma}|^2 
                  -\sum\limits_{\sigma}  |\sfac_\sigma(k)|^2 
                   \sum\limits_{\lambda} |\Meu^{\rm Born}_\lambda|^2
\end{eqnarray}
in the multiphoton distributions are all
in terms of $\sum\limits_{spin} |...|^2 $!!
We denoted: $\lambda$ = fermion helicities and $\sigma$ = photon helicity.

The above is to be contrasted with the analogous \Order{\alpha^1} case of CEEX
\begin{equation}
\sigma = \sum\limits_{n=0}^\infty\;
          \int\limits_{{ m_\gamma}} d\Phi_{n+2}
         \sum\limits_{\lambda,\sigma_1,...,\sigma_n}
          |e^{B({m_\gamma})}
          \Meu^{\lambda}_{n,\sigma_1,...,\sigma_n}(k_1,...,k_n)|^2,
\end{equation}
where the differential distributions
for $n_\gamma=0,1,2$ photons are the following:
\begin{eqnarray}
   \Meu_{0}^{\lambda} &=& \hbeta_0^\lambda,\quad \lambda={\rm fermion\;  helicities}, \nonumber \\
   \Meu^\lambda_{1,\sigma_1}(k_1) 
             &=& \hbeta^\lambda_0 \sfac_{\sigma_1}(k_1)
               +\hbeta^\lambda_{1,\sigma_1}(k_1), \nonumber \\
   \Meu^\lambda_{2,\sigma_1,\sigma_2}(k_1,k_2) 
              &=& \hbeta^\lambda_0 \sfac_{\sigma_1}(k_1) \sfac_{\sigma_2}(k_2)
       +\hbeta^\lambda_{1,\sigma_1}(k_1)\sfac_{\sigma_2}(k_2) 
               +\hbeta^\lambda_{1,\sigma_2}(k_2)\sfac_{\sigma_1}(k_1)
\end{eqnarray}
and the IR-finite building blocks are
\begin{eqnarray}
   &\hbeta^\lambda_0 = \big(e^{-B} \Meu^{\rm Born+Virt.}_{\lambda}\big)\big|_{{\cal O}(\alpha^1)}, \nonumber \\
   &\hbeta^{\lambda}_{1,\sigma}(k)=\Meu^\lambda_{1,\sigma}(k) - \hbeta^\lambda_0 \sfac_{\sigma}(k).
\end{eqnarray}
As shown explicitly, this time
everything is in terms of $\Meu$-spin-amplitudes!
This is the basic difference between EEX/YFS and CEEX.
The complete expressions for
spin amplitudes with CEEX exponentiation, for any number of photons,
are shown in ref.~\cite{ceex1:1999} for the \Order{\alpha^1} case
and in ref.~\cite{ceex2:2000} for the \Order{\alpha^2} case.

\begin{figure}[!ht]
\centering
\setlength{\unitlength}{0.1mm}
\begin{picture}(800,800)
\put( 0, 0){\makebox(0,0)[lb]{\epsfig{file=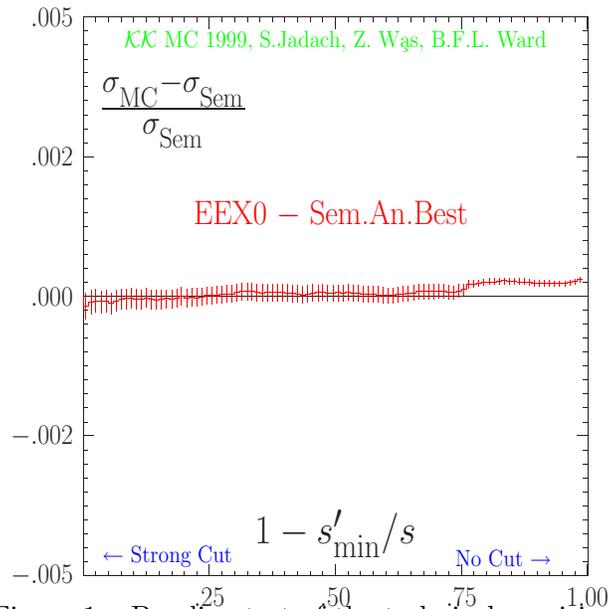,width=80mm,height=80mm}}}
\end{picture}
\vspace{-6mm}
\caption{
Baseline test of the technical precision.
}
\label{fig:technpr}
\end{figure}

\section{Monte Carlo numerical results}
The \Order{\alpha^2} CEEX-style matrix element is implemented in \KK MC
which simulates the production of muon, tau and quark pairs.
Electrons (Bhabha scattering) and neutrino channels are not available.
The program includes for the optional use the older, EEX-style matrix element.
It is then functionally similar to KORALZ~\cite{koralz4:1994} 
and the older KORALB~\cite{koralb2:1995} programs.
In Table~\ref{tab:compare} we provide the complete comparison of the features of \KK MC
and the older programs.

\begin{figure*}[!ht]
\centering
\setlength{\unitlength}{0.1mm}
\begin{picture}(1600,800)
\put( 0, 0){\makebox(0,0)[lb]{\epsfig{file=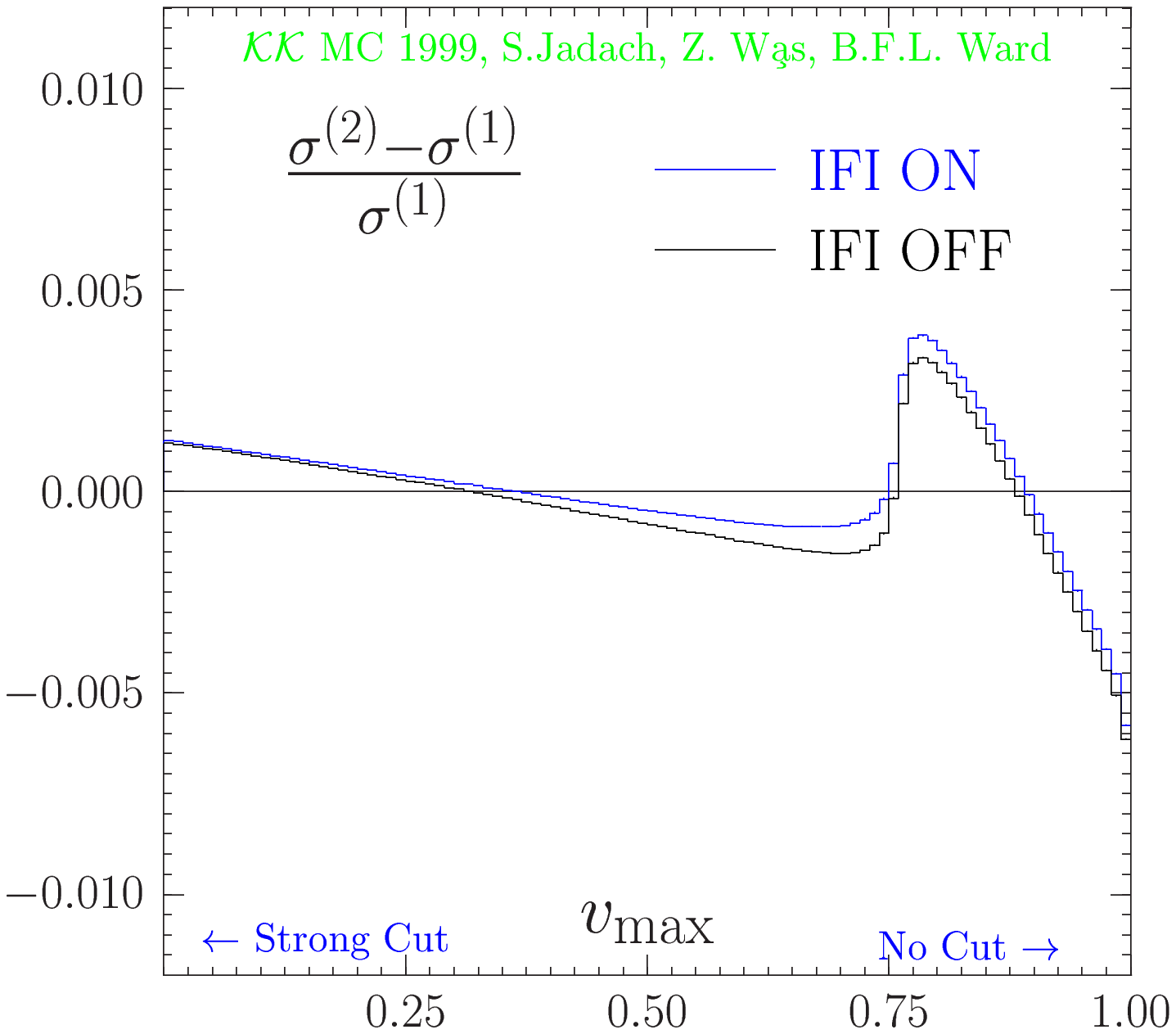,width=80mm,height=80mm}}}
\put( 800, 0){\makebox(0,0)[lb]{\epsfig{file=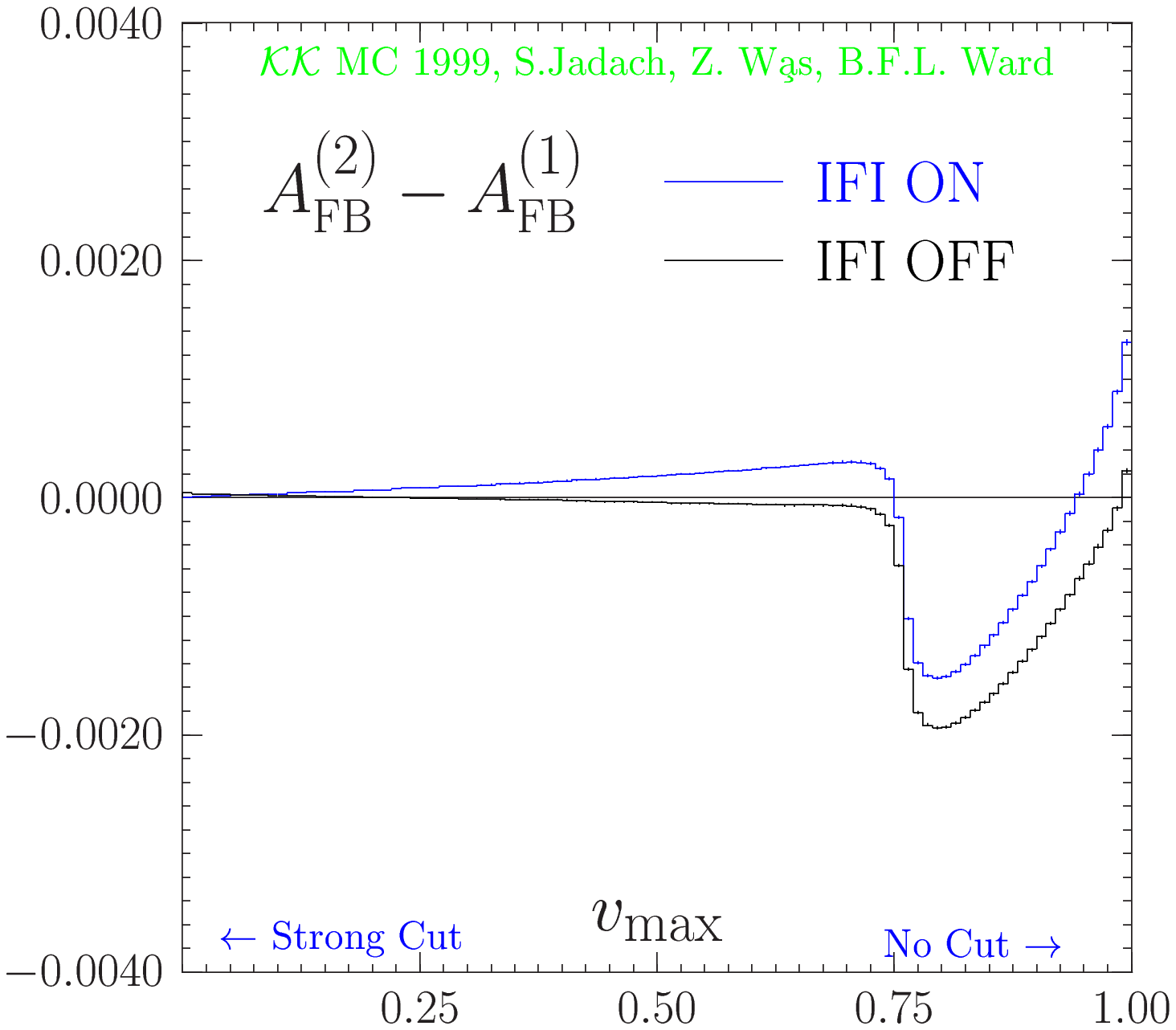,width=80mm,height=81mm}}}
\end{picture}
\vspace{-8mm}
\caption{ Test of the technical precision of \KK MC.
}
\label{fig:physpr}
\end{figure*}

\subsection{Technical precision}
For the new MC program of the high complexity like \KK MC it is important
to check very precisely the overall normalisation.
This is the cornerstone of the evaluation of the {\em technical precision}
of the program, especially for  \KK MC which is aimed at the end of testing
at the total precision of 0.1\%.
In Fig.~\ref{fig:technpr} we present the comparison of the \KK MC with simple
semi-analytical integration for the total cross section, as a function of
the minimum mass $\sqrt{s'_{\min}}$ of the final muon pair.
It is done for muon-pair final state at $\sqrt{s}=200GeV$.
For $\sqrt{s'_{\min}}\to \sqrt{s}$, when emission of hard photons is suppressed,
there is an agreement $<0.02\%$ between \KK MC and the analytical calculation.
For $\sqrt{s'_{\min}}<M_Z$ the on-shell Z-boson production
due to emission of the hard initial state radiation (ISR),
the so called Z radiative return (ZRR), is allowed kinematically.
Even in this case (more sensitive to higher orders) the agreement $<0.02\%$ is reached.
For the above exercise we used the simplified \Order{\alpha^0} CEEX matrix element,
because in this case the precise phase-space analytical integration is relatively easy.
\subsection{Physical precision}
The equally important component of the overall error is the physical error
which we estimate conservatively as the half of the difference
\Order{\alpha^2}$-$\Order{\alpha^1}.
In Fig.~\ref{fig:physpr} we show  the corresponding result for the total
cross section and charge asymmetry for $\sqrt{s}=189GeV$ as a function
of the cut on energies of all photons 
($s'_{\min}<s$ limits the total photon energy.)
We obtain in this way the estimate $0.2\%$ for the physical precision 
of the total cross section and $0.1\%$ for the charge asymmetry.
Both plots in Fig.~\ref{fig:physpr} show as expected a
strong variation at the position of the ZRR.
The quoted precision is good enough for the LEP2 combined data.

\begin{figure*}[!ht]
\centering
\setlength{\unitlength}{0.1mm}
\begin{picture}(1600,800)
\put( 0, 0){\makebox(0,0)[lb]{\epsfig{file=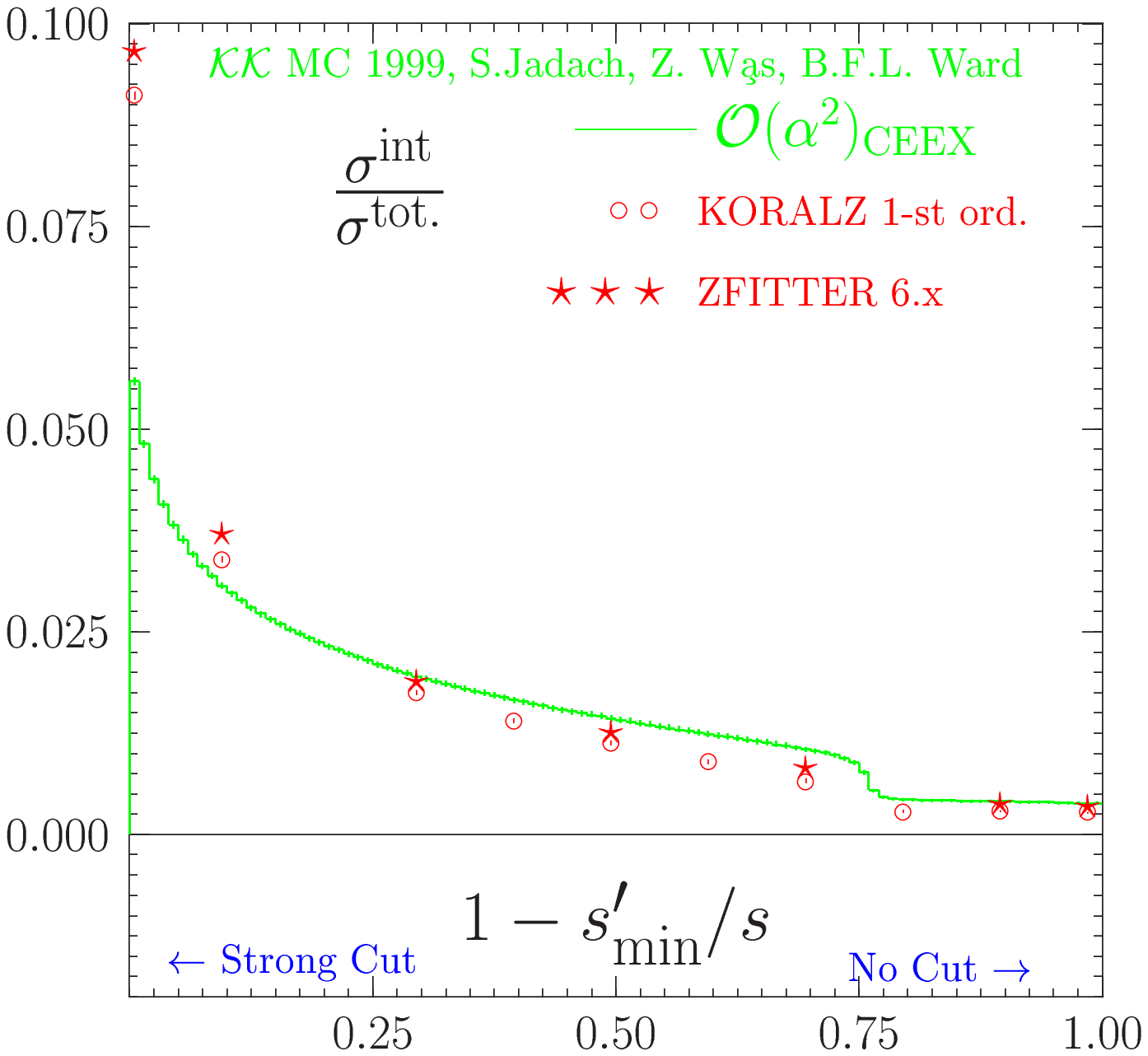,width=80mm,height=80mm}}}
\put( 800, 0){\makebox(0,0)[lb]{\epsfig{file=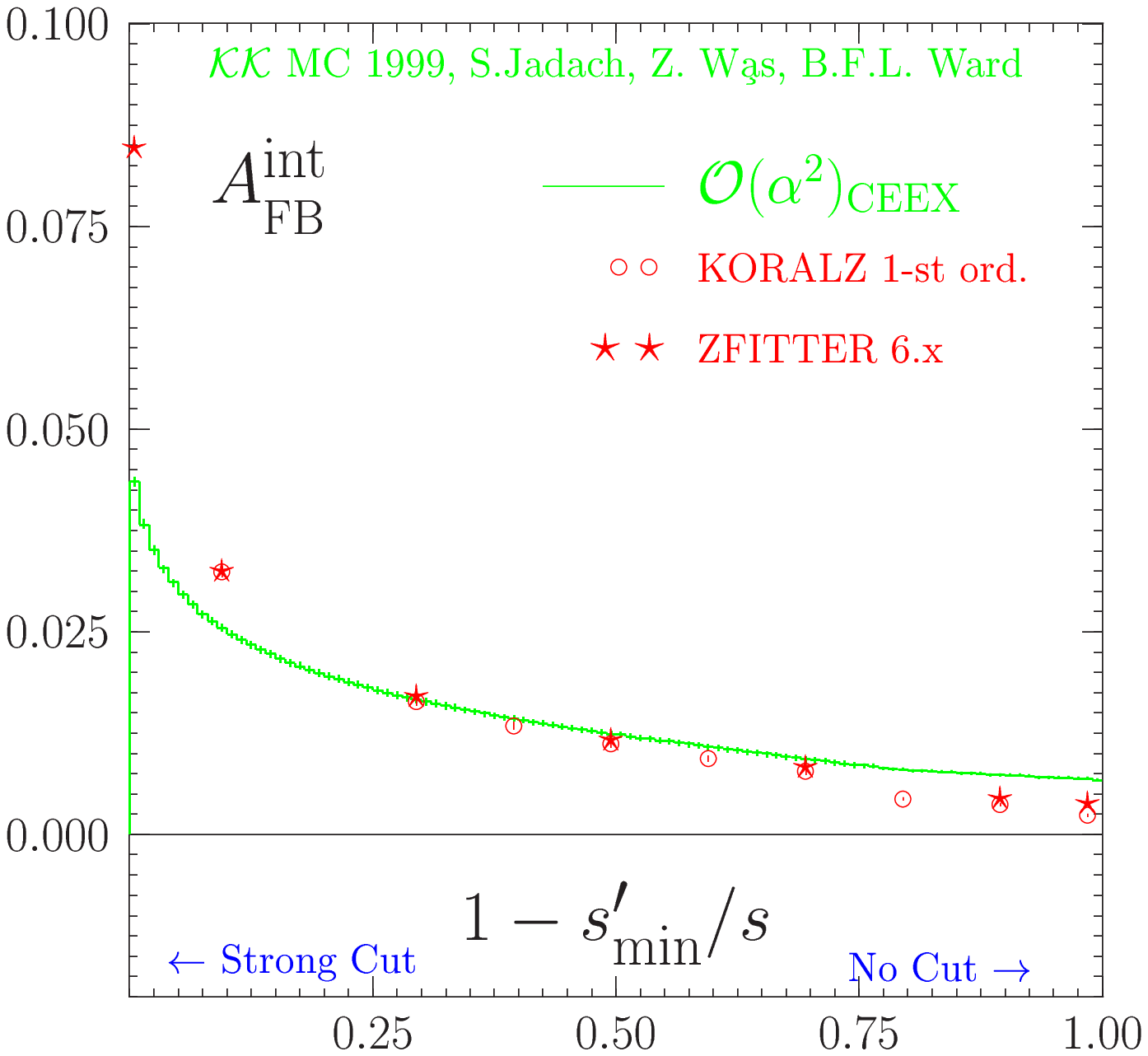,width=80mm,height=81mm}}}
\end{picture}
\vspace{-8mm}
\caption{
 The effect of the initial-final state QED interference in total cross-section
 and charge asymmetry.
}
\label{fig:IFI}
\end{figure*}

\subsection{Initial-final state QED interference}
One important benefit from CEEX with respect to the older EEX
is the inclusion of the Initial-Final state QED Interference (IFI).
The effect of the IFI is comparable with the precision of the LEP2 combined data
and should be under good control.
Results of our analysis of the size of the IFI at LEP2 energies ($\sqrt{s}=189GeV$)
are shown in Fig.~\ref{fig:IFI}.
In this figure we compare the CEEX result of \KK MC first of
all with the result of KORALZ which is
run in the \Order{\alpha^1} mode without exponentiation
(The IFI is neglected for KORALZ with the EEX matrix element.)
The \Order{\alpha^1} IFI contribution from  KORALZ was extensively cross-checked
in the past with the dedicated semi-analytical calculations~\cite{afb-prd:1991};
it is therefore a good reference and starting point.
As we see the IFI contribution of CEEX differs slightly from the pure \Order{\alpha^1}
result. 
It is related to exponentiation which makes the angular dependence (in the muon scattering angle)
of the IFI contribution less sharp and it is also due to the convolution of 
the IFI with the \Order{\alpha^2} ISR.
The expected modification of the interference correction due to higher
orders is about 20\% for the cross section and asymmetry, if the ZRR 
is excluded,
(the size of the ISR correction in the  cross section)
and it is indeed of this size.
Apparently, this principle works also in the case of ZRR included, 
remembering that in this case the ISR correction is 100\% or more.
However, we feel that this case requires further study.
We have also included results of the semianalytical program
ZFITTER~\cite{zfitter6:1999} in our plots%
\footnote{ We would like to thank D. Bardin for providing us results from ZFITTER.}.
They agree well with the \Order{\alpha^1} IFI of KORALZ.
This is expected because they are without exponentiation.

\section{Outlook and summary}
The most important new features in the present CEEX are the
ISR-FSR interference, the second-order subleading corrections, the exact matrix
element for two hard photons, and the full density matrix treatment for the spin states of initial and final state fermions\footnote{The recent presentation of the
$\tau$ lepton decay library TAUOLA can be found in ref \cite{Was:2000st}.}.
This makes CEEX already a unique source of SM predictions for the LEP2 physics program
and for the LC physics program.
Note that for these the electroweak correction library has to be reexamined at LC energies.
The most important omission in the present version is the lack of neutrino and electron
channels.
Let us stress that the present program is an excellent starting platform
for the construction of the second-order Bhabha MC generator based on CEEX exponentiation.
We hope to be able to include the Bhabha and neutrino channels soon, 
possibly in the next version\footnote{At the time of the completion of the 
conference contribution, the program version including the neutrino channel can be obtained
upon individual request only. It is still at the stage of the pre-release tests.}.
The other important directions for the development are 
the inclusion of the exact matrix element for three hard photons, together
with virtual corrections up to \Order{\alpha^3L^3} and the
emission of the light fermion pairs.
The inclusion of the $W^+W^-$ and $t\bar{t}$ final states is still in a farther perspective.



\end{document}

%% file: econfmacros2.tex
\def\beq{\begin{equation}}
\def\eeq#1{\label{#1}\end{equation}}
\def\eeqn{\end{equation}}

\newenvironment{Eqnarray}%
   {\arraycolsep 0.14em\begin{eqnarray}}{\end{eqnarray}}
\def\beqa{\begin{Eqnarray}}
\def\eeqa#1{\label{#1}\end{Eqnarray}}
\def\eeqan{\end{Eqnarray}}

\let\bar=\overbar

\def\Dslash{\not{\hbox{\kern-4pt $D$}}}
\def\dslash{\not{\hbox{\kern-2pt $\del$}}}

\def\msb{{\bar{\ssstyle M \kern -1pt S}}}

\def\lsim{\mathrel{\raise.3ex\hbox{$<$\kern-.75em\lower1ex\hbox{$\sim$}}}}
\def\gsim{\mathrel{\raise.3ex\hbox{$>$\kern-.75em\lower1ex\hbox{$\sim$}}}}